\documentclass[prl,aps,twocolumn,showpacs,amsmath,amssymb,superscriptaddress]{revtex4}
\usepackage{graphicx}
\usepackage{dcolumn}
\usepackage{bm}
\usepackage{subfigure}
\usepackage{setspace}

\begin{document}

\title{Quantum Many-Body Culling}

\author{A.M. Dudarev}
\affiliation{Max-Planck-Institut f\"ur Physik komplexer Systeme, N\"othnitzer Str. 38, 01187 Dresden, Germany}
\affiliation{Department of Physics, The University of Texas,
Austin, Texas 78712-1081} \affiliation{Center for Nonlinear
Dynamics, The University of Texas, Austin, Texas 78712-1081}
\author{M.G. Raizen}
\affiliation{Department of Physics, The University of Texas,
Austin, Texas 78712-1081} \affiliation{Center for Nonlinear
Dynamics, The University of Texas, Austin, Texas 78712-1081}
\author{Qian Niu}
\affiliation{Department of Physics, The University of Texas,
Austin, Texas 78712-1081}

\date{\today}

\begin{abstract}
We propose a method to produce a definite number of ground-state atoms by adiabatic reduction of the depth of a potential well that confines a degenerate Bose gas with repulsive interactions.  Using a variety of methods, we map out the maximum number of particles that can be supported by the well as a function of the well depth and interaction strength, covering the limiting case of a Tonks gas as well as the mean-field regime.   We also estimate the time scales for adiabaticity and discuss the recent observation of atomic number squeezing (Chuu et al., Phys. Rev. Lett. {\bf 95}, 260403 (2005)).
\end{abstract}


\maketitle

The controlled generation of many-body atomic number states has been a long-standing goal in physics and success would open a door to a controlled study of entanglement~\cite{ent1,ent2,ent3,ent4}, few-body tunneling~\cite{tun} and could also find important applications in quantum computing.  One avenue towards this goal is the Mott insulator state where single-atom or multi-atom number states are predicted~\cite{mi1,mi2}.  Most experiments to date have used  optical lattices~\cite{mi_exp} where direct access to individual sites has not been accomplished and appears very difficult. A completely different approach was used in a recent experiment.  The resulting atomic number squeezing was directly measured by atom counting, and a number state was inferred by accounting for known noise sources~\cite{chuu05}. In this Letter we analyze this new approach.

The basic idea is to confine a degenerate Bose gas in an optical box with finite barrier height that can be controlled.  The repulsive interaction between the atoms means that a finite box can only contain a maximum number of atoms.  As the barrier height is slowly reduced, atoms must leave, and the final number will be completely determined by the stopping point of the barrier.  Since the confinement and the barrier in Ref.~\cite{chuu05} were realized by means of dipole optical traps we call this process ``laser culling of atoms'' although the same principle could also be implemented in other types of traps.  The main theoretical questions that need to be addressed are how the maximum number depends on the potential parameter and interaction strength and how slow the potential should be changed in order to avoid excitations within the box.

We address these questions by calculating and analyzing the energy levels of $N$-particle ground states as well as excitations that are bound to the potential well.  This is in general a very difficult task because of the interactions between the atoms.   Therefore, we consider a simple model of one dimensional (1D) bosons with contact interactions in a square well potential.   This model captures the essential features of the experimental system used recently to produce atomic number squeezing \cite {chuu05}. 

The letter is structured as following. First, we describe the model. Then we outline the idea of atomic culling in the limiting case of impenetrable bosons, which is followed by the discussion of another limit where modified mean-field picture is relevant. The diffusion Monte Carlo and direct diagonalization of the Hamiltonian bridge these two regions. Finally, we discuss the criteria for adiabaticity of the process and effect of the initial temperature.

As usual, the interaction strength $g$, the coefficient of the $\delta$-function interaction potential, depends on the three dimensional s-wave scattering length $a_s$ as well as the widths of the transverse wave functions $a_\bot$: $g = 2 \hbar^2 a_s / m a_\bot^2$ when $a_\bot \gg a_s$~\cite{olshanii98}. One can therefore vary $g$ by changing the transverse confinement potential.   The potential well depth $V_0$ is controlled by the laser intensity for the barriers.    The width $L$ of the potential well is another parameter that can be adjusted over a wide region.   Our model is made dimensionless by using the convention that the Planck constant $\hbar$, atomic mass $m$, and the well width $L$ are all unity.   Thus, there are two dimensionless parameters in our model, the well depth in units of $(\hbar/L)^2/m$ and the interaction strength in units of  $\hbar^2/m L$.

%Repulsive interaction is absolutely needed to achieve number selection, because otherwise all the boson atoms can fit into the ground state in the well no matter how many they are.    In general, strong interaction makes bigger energy gaps between different number states or for excitations within the same particle number.     An ideal limit for our purpose is therefore the Tonks gas, in which the atoms become impenetrable to each other.

% \section{Introduction}

%Paper by Holland, paper by Kolomeyski, experimental paper, tweezer paper, multiparticle interferometry paper and all significant papers from there, Meschede group papers. Emphasis on 1D since the particle-particle interaction is strong and it is obtained.

%The level splitting is the largest when the typical confining length is the smallest. This can be achieved in experimentally relevant configuration of such a strong confinement in two directions that the dynamics is confined to one dimension. Such situation was recently achieved with neutral atoms in magnetic and optical dipole traps. Below we focus on this configuration as the most practically relevant.

%Below with various approaches and in various limits we consider the same system: $N$ one-dimensional bosons interacting with potential $U(x_1,x_2)=g \delta (x_1- x_2)$ located in the square well of depth $V_0$ and width $L$. The units are such that $\hbar = 1$, mass of atoms $m=1$. We also choose unit of length so that $L=1$.

% \section{Introductory description}

\begin{figure}
\includegraphics[width=7.5cm]{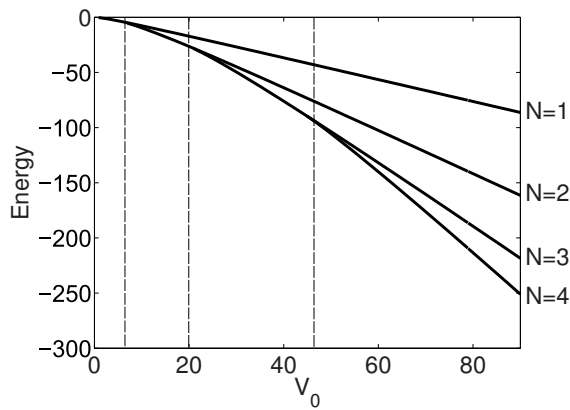}
\caption{Ground state energy of $N$ Tonks bosons in a square well. Zero of energy is at the top of the well. For $V_0$ where two levels merge larger number of particles is not supported. These values are indicated by vertical dashed lines.}\label{fig:levels_tg}
\end{figure}

% how delocalization happence
To explain how quantum culling happens we consider $N$ impenetrable bosons in a 1D well. In this situation the problem is mapped exactly to $N$ non-interacting fermions in the same potential~\cite{girardeau60}. The eigenstates $\Psi (x_1, \cdots x_N)$ of the problem are given by properly symmetrized set of single particle eigenstates $\phi_j(x)$
\begin{equation}
	\Psi(x_1, \cdots x_N) = \hat S \phi_{j_1}(x_1) \cdots \phi_{j_N}(x_N),
	\label{eq:product}
\end{equation}
which in this situation for bosons is given by the absolute value of the Slater determinant. The ground state corresponds to indices $j$ arranged from $1$ to $N$.  Since the finite 1D well supports $N$ bound single particle state only for 
\begin{equation}
        V_{0,N} > (\pi (N-1)/L)^2/2,
	\label{eq:tg}
\end{equation}
only then $N$ impenetrable bosons can be trapped in it. As the well depth decreases to such a value that the most energetic single particle state delocalizes, only $N-1$ particles remain trapped in the well (see Fig.~\ref{fig:levels_tg}). If initially only $N-1$ atoms were trapped, they remain trapped. Hence if initially there was uncertainty in the number of atoms in the trap ($N$ or $N-1$) it is reduced. 

\begin{figure}[t]
\includegraphics[width=7.5cm]{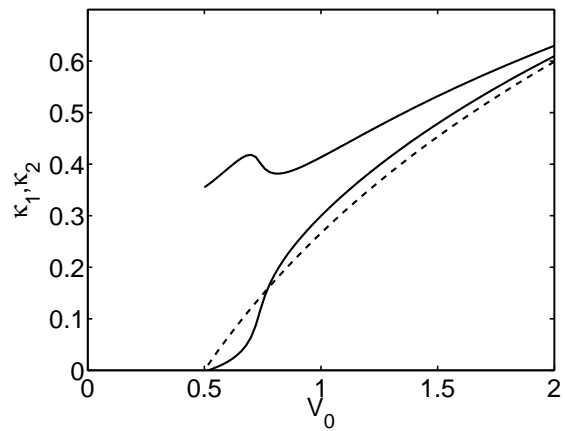}
\caption{Variational parameters $\kappa_1$ and $\kappa_2$ that minimize the total energy for $N=3$ and $g=1$. Dashed line shows the variational parameter for a wave function that uses only one variational parameter.}\label{fig:kappa}
\end{figure}

In the opposite limit of weak interaction, one may invoke mean-field approximations.    The simplest approach assumes that all the $N$ particles are in the same single particle state, with its wave function determined variationally by the standard Gross-Pitaevskii (GP) equation~\cite{pitaevskii61,gross61}.    This is only good for the ground state with all the particles deeply bound in the well.   Near ``ionization'' threshold, only one particle should become weakly bound with long tails of the corresponding orbital reaching outside the well, while all the others remain tightly bound.    This situation is better described by a Hartree-Fock wave function with $N-1$ particles in a state $\phi_1$ and one particle in another state $\phi_2$~\cite{cederbaum03,masiello05}.
In order to minimize the total energy, $\phi_1$ should be symmetric and nodeless. For bosons, $\phi_2$ does not need to be orthogonal to $\phi_1$, and should in fact also be symmetric and nodeless, because the $N$-body wave function should be symmetric and nodeless with respect to any particle coordinate in the ground state.  In Fig.~\ref{fig:kappa}, we show the result of variational minimization of the total energy for the case of $N=3$, and $g=1$, using variational functions of the form 
\begin{equation}
	\phi_i(x) = \sqrt{\kappa_i} \exp (- \kappa_i \left| x \right|) {\rm~ ~} (i=1,2).
\end{equation}
We minimize the energy varying parameters $\kappa_1$ and $\kappa_2$ in the symmetrized wavefunction~(\ref{eq:product}) for a given strength of interaction $g$ and depth of the potential $V_0$. We allow only a single particle to populate the second orbital. Non-vanishing overlap between the orbitals must be taken into account. For a deep well $\kappa_i$ are comparable, which indicates a condensate state. As the depth of the well is reduced $\kappa_2$ tends to zero, which corresponds to delocalization of one particle. In comparison we show the result of minimization of total energy when only one variational parameter is used, then vanishing of the variational parameter corresponds to delocalization of all the particles. For all the depths of the potential the energy is smaller when two-orbital wave function is used.

% inadequacy of GP equation to describe evaporation
In the limit of large number of particles and weak interaction the Thomas-Fermi approximation can be used. The approximation is valid when $N g \gg 1/L$, where $N$ is number of particles. Delocalization of the wave function in the standard GP equation in Thomas-Fermi limit can be expected when the chemical potential becomes equal to the level of the top of the well ($\mu = 0$). This corresponds to a depth of the potential $V_0 < g (N-1) / L$.  When two asymmetric orbitals $\phi_1(x)$ and $\phi_2(x)$ in the mean-field treatment~\cite{cederbaum03} are considered, minimization of energy functional leads to two coupled nonlinear equations
\begin{equation}
	\begin{array}{l}
		\left( {h(x) + g(N  - 2)\left| {\phi _1 } \right|^2  + 2 g\left| {\phi _2 } \right|^2 } \right)\phi _1  = \mu _1 \phi _1, \\ 
		\left( {h(x) + 2 (N - 1) g\left| {\phi _1 } \right|^2 } \right)\phi _2  = \mu _2 \phi _2,
	\end{array}
\end{equation}
where $h(x)$ is a single particle Hamiltonian. Neglecting kinetic energy term in the Hamiltonian and considering the situation when the orbital $\phi_2$ is almost delocalized, we see that the second equation is a single particle equation with potential modified by the rest of the atoms. As a result $N$ atoms are supported in the potential when
\begin{equation}
	V_0 > 2 g (N-1) / L.
	\label{eq:tf}
\end{equation}
The energy calculated with two orbitals approach is smaller than calculated with a single orbital when $N g > L V_0$, which is consistent with the approximations.

% \section{Calculation $N$ vs $V0$, $g_{1D}$ with two limiting cases (no dynamics)}

% description of the second quantized approach

\begin{figure}[t]
\includegraphics[width=7.5cm]{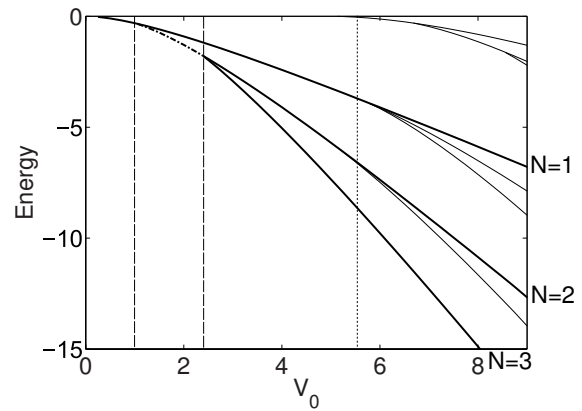}
\caption{Bound levels of bosons interacting with $g=1$. Zero of energy is
at the top of the well. Thick lines correspond to ground state of $N$
particles. Thin lines indicate bound excitation. Dashed-dotted line shows
region when $N=2$ atoms are bound and $N=3$ are not. Two vertical dashed lines indicate the regions where only $N=1,2$ particles are supported in the bound state (from left to right). Vertical dotted line shows the smallest $V_0$ for which excited bound state is supported for $N=3$.}\label{fig:levels_g}
\end{figure}

To describe the system for small values of $g$ and small number of particles we resort to direct diagonalization of the many-body Hamiltonian. We limit the well by a box with infinite walls of size $D\gg L$. A single particle wave functions $\psi_j(x)$ can be either even or odd and inside of the well are proportional to $\cos(kx)$ and $\sin(kx)$ correspondingly. Outside of the well they are given by linear combination of exponents $\exp(\pm \kappa x)$, with $\kappa = \sqrt{k^2 - 2 V_0}$, and such coefficients that the wave function is continuous at the boundary. When $\kappa$ is imaginary the single particle states are extended. To find several single particle states, the values of $k$ are adjusted so that the wave function vanishes at the boundary of the larger well. The $N$ particle basis is constructed with these states. In the basis of second quantized states with $N_j$ atoms on $j$'s single particle state, single particle energies $E_j$ contribute to diagonal terms of the Hamiltonian
\begin{equation}
	\hat H _1  = \sum\limits_j {E_j \hat n_j },
\end{equation}
while interaction gives contribution to other terms as well
	\begin{eqnarray}
		\hat H_2  = \frac{1}{2}\sum\limits_{j_1,j_2,j_3,j_4} {\hat a_{j_1 }^{\dagger} } \hat a_{j_2 }^{\dagger} \hat a_{j_4 } \hat a_{j_3 } \left\langle {j_1 j_2 } \right|U\left| {j_3 j_4 } \right\rangle, \\
		\left\langle {j_1 j_2 } \right|U\left| {j_3 j_4 } \right\rangle  = \int {dx\psi^* _{j_1} \psi^* _{j_2} \psi _{j_3} \psi _{j_4} }.
	\end{eqnarray}
	We diagonalize the full Hamiltonian $\hat H = \hat H_1 + \hat H_2$. The bound states for $N=1,2,3$ and $g=1$ are shown in Fig.~\ref{fig:levels_g} for different $V_0$. 

\begin{figure}[t]
\includegraphics[width=7.5cm]{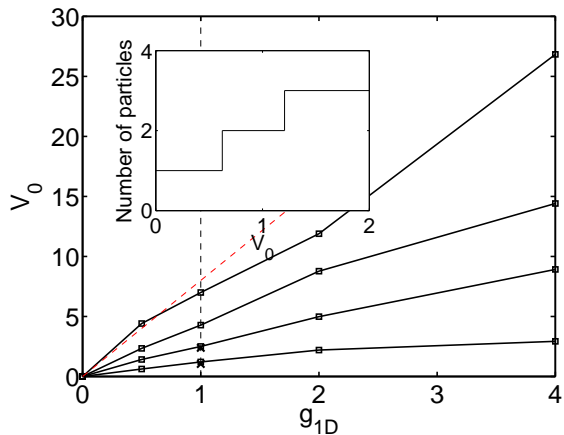}
\caption{Parameter space for $N$ atoms to be bound by the well. Regions from bottom to top $N=1,2,3,4$, and $N\ge5$. Inset shows change in number of atoms along vertical dashed line. The squares are obtained with diffusion Monte Carlo method, the lines are guides for an eye. Two crosses for $g=1$ are obtained with diagonalization of full Hamiltonian. The inclined dashed line is the transition between $N=4$  and $N=5$ atoms in Thomas-Fermi limit, Eq.~(\ref{eq:tf}).}\label{fig:parameter_space}
\end{figure}

As an alternative method to calculating the ground state of $N$ bosons, we use the diffusion Monte Carlo approach~\cite{astrakharchik04}. Previously this method has been successfully applied to calculate the ground state of bosons with arbitrary interaction. With this approach we also use a box of the size $D \gg L$. From the depths of the well $V_0$ such that $N$ particles have smaller energy than $N-1$ particles we extrapolate smaller depth of the well and determine where the bound state of $N$ particles is not bound anymore. Such a procedure leads to errors due to the finite time of the diffusion Monte Carlo evolution, finite size of larger box and extrapolation. In Fig.~\ref{fig:parameter_space} we show regions where $N$ particles can be supported. The errors are estimated to be comparable to the size of the symbols. Transition from $N=1$ to $N=2$, and from $N=2$ to $N=3$ for $g=1$ agrees quite well with diagonalization of the second quantized Hamiltonian. Transition from $N=4$ to $N=5$ is close to what is predicted by Thomas-Fermi formula Eq.~(\ref{eq:tf}). 

% on adiabaticity limit

We now discuss how to achieve the atomic number states that we have studied above.  We assume that our initial state is one with an unknown number of atoms but at essentially zero temperature.  If the initial temperature is finite, a cooling procedure needs to be applied, which can be done by standard evaporation techniques.  Therefore, one can prepare an initial state corresponding to one of the ground states that we have calculated above.  However, the atomic number may not be the maximum allowed for the potential well, so that it is impossible to determine the initial atomic number from the depth of the potential well.  One can then lower the barriers slowly until a threshold where it is no longer possible to hold the initial number of atoms. Then one atom will leave the well and there will be one less atom in the well.  From this point on, as the potential barriers continue to reduce, there will be a one to one correspondence between the atomic number in the well and the intervals between the threshold values of the potential well.  

\begin{figure}
\includegraphics[width=7.5cm]{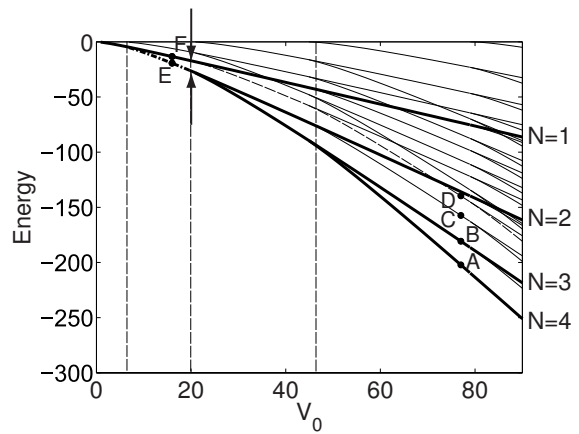}
\caption{Bound levels of impenetrable bosons in a square well in dependence on the depth of the well. Zero of energy is at the top of the well. Thick lines correspond to ground state of $N$ particles (same as Fig.~\ref{fig:levels_tg}). Thin lines indicate excitations. Dashed-dotted line shows
region when $N=2$ atoms are bound and $N=3$ are not. Two arrows show the gap relevant during the final stage. If system starts at points A, B, or C it adiabatically evolves to point E ($N=2$). Starting at point D it evolves to point F ($N=1$). A dashed level indicates excited state of $N=2,3,$ and $4$  configuration above which the system adiabatically evolves into $N=1$ state. Vertical dashed lines show the regions where only $N=1,2,3$ particles are supported (from left to right).}\label{fig:levels_tg_2}
\end{figure}

The discussion above assumes that initially system is not excited and that
the excitations are not created during the process. There are two types of
excitations: those when number of bound atoms is fixed and those when some
atoms escape. For example, in Fig.~\ref{fig:levels_tg_2} the transition
between points B and C corresponds to the first type when three bound atoms
remain bound. On another hand, the transition indicated with arrows on the same figure corresponds to the second type when after excitation of two bound atoms one becomes unbound. All transitions between bold lines are of this type. When
one wants to control atoms with a single atom precision these excitations
are posing the main fundamental limitation. When bound state of $N+1$ atoms
becomes unsupported the state with $N$ atoms has only one bound state.
Hence, only unbounding excitation are possible at the final stage.
To illustrate this with a specific example, examine Fig.~\ref{fig:levels_g}
and~\ref{fig:levels_tg_2}. There the level with $N=2$ atoms is highlighted with
dashed-dotted line when
$N=3$ has just become unsupported. Above it there are no excited states
with $N=2$. Also if one wants to obtain two atoms he better stop at the
right side of the indicated interval, because on the right side the
relevant excitation gap vanishes. First, we discuss the heating excitations (when number
of atoms is preserved) in the connection with recent experiments, and
later, consider the unbounding excitations during the final stage.

During the evolution the potential barrier must vary sufficiently slowly in
order to keep the system always in the ground state for each given number
of atoms. The Planck constant divided by the excitation energy gaps give a
time scale during which the wave function cannot change significantly for
the evolution to remain adiabatic. For this problem the wave function
undergoes significant changes in the range of the potential from one atom
delocalization to another. 

In the recently reported experiment~\cite{chuu05}, sub-Poissonian number statistics were directly observed by atom counting.  The measured variances were nearly a factor of two below the shot-noise and could be attributed to known sources of technical noise.  These observations were therefore consistent with the production of number states.
The change in atom number uncertainty was studied for different rates with
which the depth of the potential is changed. Increase in uncertainty for
larger rates can be attributed to generation of bounded excitation. We took
the shape of the potential change as in the paper, $V_{\rm depth}(t)=V_0 e
^ {- t/\tau}$, and assumed the dependence of the number of atoms on the depth of the potential to be in the form $N\propto V_{\rm depth} ^\alpha$, with power $\alpha$ of order of one. We use GP equation and estimate the time scale of the excitation $t_{\rm gap}$ from the energy of the Bogoliubov excitation of the longest wavelength~\cite{pethick}
$\omega = (\rho U_0/m)^{1/2} \pi / L_x$, where $U_0 = 4 \pi \hbar a/m$, $a$
is the scattering length, $m$ is the atom mass, $\rho$ is the density,
$L_x$ extension of the wave function in the non-confined direction.  We
would like to emphasize that the numbers below may be considered only as an
order of magnitude estimate since as discussed above the delocalization
depth may be different by some factor from what expected with GP equation.
We estimated the average density by taking the parameters of the trap and
numerically finding the ground state of the 3D GP equation, changing the
degree of nonlinearity to assure confinement in the well. For the
parameters of the experiment for $N_{\rm final}$ we find $t_{\rm gap}
\approx 5$ ms, and the evolution is adiabatic when $\tau \gg 60$ ms, consistent within an order of magnitude observed in the experiment. If one changes the well faster, excitations to the levels corresponding at the end to smaller number of atoms are unavoidable, resulting in larger spread in the distribution of number of atoms after the potential change as observed experimentally.

As discussed above, the unbounding excitations put the most fundamental
restriction when control with precision of one atom is required. We consider unbounding excitations during the final stage in two limits. In
the limit of strong interaction (Tonks gas), the system behaves like a
non-interacting fermion gas, and the excitations correspond to promoting
the atoms from the filled single particle levels to the empty ones. In
Fig.~\ref{fig:levels_tg_2}, we plot the energies of the lowest excited
state together with the ground state energy for each $N$. During the final
stage the relevant excitation energy gap is energy difference between $N$-
and $(N-1)$-particle states when $N+1$ particle have just stopped being
supported. We illustrate this gap for $N=2$ particles on
Fig.~\ref{fig:levels_tg_2}. It can be calculated as the difference of two
solutions of a transcendental equation. In the limit of large $N$ it is $E
_{\rm gap} = \pi ^2 (N+1/2)$. We would like to note that for $N=1,2$ the
similar linear dependence $E_ {\rm gap} = A N + B$ holds except the slope
$A$ is approximately two times smaller.  The depth of the well below which
$N$ particle cannot be bound is given by Eq.~\ref{eq:tg}. As discussed
above, the wave function varies significantly in the range of depth
$V_{0,N+1} - V_{0,N} = \pi^2 (N-1/2)/L^2$. The condition on potential depth
rate of change, $r=V'_0 (t)$, to be adiabatic is then (Planck constant here
is unity as above)
\begin{equation}
        \frac{r}{E_{\rm gap}} \ll V_{0,N+1} - V_{0,N}.
\end{equation}
It follows that for large $N$ the maximum rate is proportional to $N^2$. This can be understood from the Fig.~\ref{fig:levels_tg_2}: both gaps and gap intervals of $V_{0,N}$ are proportional to $N$. As a result is in this limit when one goes to smaller and smaller number of atoms it becomes more and more difficult to remain adiabatic. In the case of $N=1$ the maximum rate (in dimensionless units) is approximately unity (it does not vanish, because both gap and the depth intervals do not vanish). 

In the opposite limit of weak interaction, where the mean field picture is
applicable, the energy gap is given by the energy difference between $N$
and $N+1$ particles when the latter stops being supported. For large $N$ it
is approximately $E _{\rm gap} = 3 g N^2 / 2$. The range of $V_0$ for which
$N$ particles are supported in this limit is $N$ independent $V_{0,N+1} -
V_{0,N} = 2g/L$ (see Eq.~\ref{eq:tf}). Hence in this limit maximum rate $r \propto g^2 N^2$. It also becomes more and more challenging to obtain smaller number of atoms. The quadratic dependence on $g$ shows that it is advantageous for the experimentalists to increase the effective interaction to make the maximum allowed rate larger.

Finally, we would like to mention effect of initial excitation for the process. As seen of Fig.~\ref{fig:levels_g} and Fig.~\ref{fig:levels_tg_2} one may start from thin lines corresponding to excitations in $N>1$ systems and read the final state with $N=1$. In general, in both limits, if the final goal is to get $N$ bound atoms and one starts with state with $M$ atoms, $M-N$ lowest many-body bound excitations are allowed. This means that for $M\gg N$ one does not even have to start in degeneracy limit for the process to be possible.

In conclusion, we have considered a system of $N$ interacting bosons in a
finite well. In such system a standard mean-field approach is not
applicable. We have used diffusion Monte Carlo approach to obtain the
parameters for $N$ atoms to remain bound. The calculations agree with
direct diagonalization for small number of atoms and analytical formulas in
the limiting cases. As the depth of the well changes too fast excitation to
levels that evolve into smaller number of particles can occur. We have
estimated the critical rate for the recent experiments, which agreed within
an order of magnitude. The limitations due to excitation during the last
stage when precision within a single atom is required are discussed.

%\section{Dynamics in mean-field regime (connection to experiments)}

%Absorbing boundary condition with GP. Order of magnitude for critical rate estimation. Things it may depend on: velocity of the sound, size of the condensate. Consider two regimes: dynamically changing depth, dynamically changing chemical potential.

%\section{Dynamics in TG regime}

%Strong correlation. Connection to fermions. The best one may hope with boson. Point out that clearly reducing chemical potential is more advantageous than dynamically changing depth of the potential.

%\section{Conclusion and acknoledgements}

AD acknowledges useful discussions with Chih-Sung Chuu, Sungyun Kim, Joachim Brandt and Grigory Astrakharchik.
MGR acknowledges support from NSF, the R. A. Welch Foundation, and the S. W. Richardson Foundation and the US Office of Naval Research, Quantum Optics Initiative,
Grant N0014-04-1-0336.


\begin{thebibliography}{99}
	\bibitem{ent1} 
		D. Jaksch, H.-J. Briegel, J. I. Cirac, C. W. Gardiner, and P. Zoller,
		Phys. Rev. Lett. {\bf 82}, 1975  (1999).
	\bibitem{ent2} 
		T. Calarco, E. A. Hinds, D. Jaksch, J. Schmiedmayer, J. I. Cirac, and P. Zoller,
		Phys. Rev. A {\bf 61}, 022304 (2000).
	\bibitem{ent3} E. Andersson and S. M. Barnett, Phys. Rev. A {\bf 62}, 052311 (2000).
	\bibitem{ent4}
		A.M. Dudarev, R.B. Diener, B. Wu, M. G. Raizen, and Q. Niu,
		Phys. Rev. Lett. {\bf 91} 010402 (2003).
	\bibitem{tun} S. Kim, A.M. Dudarev, and J. Brand {\it to be published}.
	\bibitem{mi1} 
		M.P.A. Fisher, P.B. Weichman, G. Grinstein, and D.S. Fisher,
		Phys. Rev B, {\bf 40}, 546 (1989).
	\bibitem{mi2} 
		D. Jaksch, C. Bruder, J.I. Cirac, C.W. Gardiner, and P. Zoller,
		Phys. Rev. Lett., {\bf 81}, 3108 (1998).
	\bibitem{mi_exp} 
		M. Greiner, O. Mandel, T. Esslinger, T.W. Hansch, and I. Bloch,
		Nature, {\bf 415}, 39 (2002).
	\bibitem{chuu05} 
		C.-S. Chuu, F. Schreck, T.P. Meyrath, J.L. Hanssen, G.N. Price, and M.G. Raizen,
		Phys. Rev. Lett. {\bf 95} 260403 (2005).
	\bibitem{olshanii98} M. Olshanii, Phys. Rev. Lett. {\bf 81}, 938 (1998).
	\bibitem{girardeau60} M. Girardeau, J. Math. Phys. (N.Y.) {\bf 1}, 516 (1960).
	\bibitem{pitaevskii61} L.P. Pitaevskii, Sov. Phys. JETP {\bf 13} 451 (1961).
	\bibitem{gross61} E.P. Gross, Nuovo Cimento {\bf 20} 454 (1961) .
	\bibitem{cederbaum03} L.S. Cederbaum and A.I. Streltsov, Phys. Lett. A {\bf 318} 564 {2003}.
	\bibitem{masiello05} D. Masiello, S. B. McKagan, and W. P. Reinhardt, e-print cond-mat/0509530 (2005).
	\bibitem{astrakharchik04} G. E. Astrakharchik, Ph.D. thesis, Universita di Trento, 2004
	\bibitem{pethick} C.J. Pethick and H. Smith, {\it Bose-Einstein Condensation in dilute Gases} (Cambridge University Press, Cambridge, 2002) Ch. 7.
\end{thebibliography}
\end{document}